\documentstyle[twocolumn,aps,prl,epsfig]{revtex}
\twocolumn

\begin{document}
\draft
\title{Electromagnetically induced absorption in magneto-optically trapped atoms.}
\author{A. Lipsich, S. Barreiro, P. Valente and A. Lezama\thanks{%
E-mail: alezama@fing.edu.uy}}
\address{Instituto de F\'{i}sica, Facultad de Ingenier\'{i}a. Casilla de correo 30. \\
11000, Montevideo, Uruguay.}
\date{\today }
\maketitle

\begin{abstract}
Electromagnetically induced absorption (EIA) was observed on a sample of $%
^{85}Rb$ in a magneto-optical trap using low intensity cw copropagating pump
and probe optical fields. At moderate trapping field intensity, the EIA
spectrum is determined by the Zeeman effect produced on the atomic
ground-state by the trapping quadrupolar magnetic field. The use of EIA
spectroscopy for the magnetic field mapping of cold atomic samples is
illustrated.
\end{abstract}


\preprint{}

\section{Introduction.}

The interaction of radiation with a coherently prepared atomic medium has
attracted considerable attention in recent years \cite{SCULLYBOOK}. While
most coherent processes have been studied using three or more atomic energy
levels, it was recently demonstrated that degenerate two-level systems
constitute an interesting host for coherent spectroscopical effects \cite
{LEZAMA,LEZAMA2,AKULSHIN99}. A new coherent effect observed in degenerate
two-level systems, is the increase of the atomic absorption occurring when
the lower atomic energy level has a smaller (angular momentum) degeneracy
than the upper level and the frequencies of two mutually coherent optical
fields match the condition for Raman resonance between ground-state Zeeman
sublevels. This phenomenon, the change in sign taken apart, share several
properties with electromagnetically induced transparency (EIT) \cite{HARRIS}
and was designated electromagnetically induced absorption (EIA) \cite
{LEZAMA,LEZAMA2}. The spectral properties of EIA depending on field
intensity, magnetic field and optical fields polarizations were recently
analyzed \cite{LEZAMA3}. In particular, the sensitive dependence of the EIA
spectrum on magnetic field makes it a potentially useful tool for the
characterization of the magnetic atomic environment \cite{SCULLY92}. To
date, the reported observations of EIA were carried in vapor samples. It is
worth noticing that the atomic transitions normally used for magneto-optical
trapping and cooling verify the conditions for EIA \cite{LEZAMA2}: they are
closed transitions with a larger angular momentum in the excited state than
in the ground state. This paper presents the first observation of EIA on a
cold atomic sample in a magneto optical trap (MOT) using the trapping
transition. It is shown that EIA constitute a simple and direct tool for the
inspection of the atomic density distribution with respect to the trapping
quadrupolar magnetic field \cite{SAVARD}.

We have performed EIA spectroscopy on a sample of magneto-optically cooled $%
^{85}Rb$ atoms in the presence of the trapping MOT field. In addition to the
fields necessary for the MOT operation, the sample was submitted to a fixed
frequency pump field and a tunable probe field with linear and mutually
orthogonal polarizations. Both fields were quasiresonant with the closed $%
5S_{1/2}(F=3)\rightarrow $ $5P_{3/2}(F^{\prime }=4)$ transition of $^{85}Rb$
(trapping transition). A copropagating geometry was used for the pump and
the probe waves. This geometry has the advantage of avoiding a spatial
dependence of the relative phase between the pump and the probe field that
would exist, for instance, in a counter-propagating geometry and allow more
direct comparison with theory. However, the copropagating geometry has the
disadvantage of requiring very low intensity in both the pump and probe
waves to avoid ``blowing'' the trapped atoms by the radiative force exerted
by the quasiresonant fields. Consequently, EIA was observed only in a regime
of pump and probe intensities many orders of magnitude below saturation for
which the perturbation of the MOT by the pump and probe field was observed
to be negligible. Since small field intensities were used, the corresponding
EIA resonances were small compared to the linear absorption. This required
the use of a highly sensitive detection technique in order to distinguish
the EIA\ resonances from the large linear absorption background. A
double-frequency lock-in detection was used.

\section{Setup.}

The experiments were done on a standard magneto optical trap for $^{85}Rb$
with three pairs of counterpropagating trapping beams along orthogonal
directions. The maximum available total trapping field power was $20$ $mW$
and the trapping beam cross section $1\ cm$. A repumping field of $1$ $mW$
generated by an independent laser was superimposed to the trapping beams.
The MOT used a quadrupolar magnetic field distribution with a magnetic field
gradient of $1\ G/mm$ along the symmetry axis (vertical). The MOT produced a
cold atom cloud of typical $0.7\ mm$ transverse dimension containing (at
maximum trapping field power) $10^{7}$ atoms. The trapping and pump fields
were obtained from the same (frequency stabilized) extended cavity diode
laser (master laser) using in each case the following procedure: the output
of the master laser was frequency shifted by an acousto optic modulator
(AOM) to a fixed frequency position relative to the $5S_{1/2}(F=3)%
\rightarrow $ $5P_{3/2}(F^{\prime }=4)$ atomic transition and used to
injection-lock a separate laser diode for power increase. As in \cite{LEZAMA}%
, the probe field was obtained from the pump by the use of two additional
consecutive AOMs, one of them driven by a tunable RF generator. In this way,
all three trapping, pump and probe fields were highly correlated. The pump
and probe waves with orthogonal linear polarizations are superimposed on a $%
50\ cm$ long polarization-preserving single mode optical fiber and
subsequently focussed on the cold atoms sample ($0.3\ mm$ beam waist).
Intensity was controlled with neutral density filters in order to have
approximately the same power in the pump and probe waves. At the sample,
both waves had a central intensity around $0.1\ \mu W/cm^{2}$. The pump and
the probe fields were mechanically chopped at frequencies $f_{1}$ and $f_{2}$
($1$ and $1.2\ kHz$ respectively). After the sample, a photodiode detected
the transmitted light. The photodiode output was analyzed with a lock-in
amplifier using as reference the sum frequency $f=f_{1}+f_{2}$. With this
technique, only the nonlinear component of the absorption, proportional to
the product of the pump and probe field intensities was detected. The
spectra were recorded by scanning the frequency offset between the tunable
probe and the fixed-frequency pump.

\section{Results and discussion.}

\begin{figure}[tbp]
\begin{center}
\mbox{\epsfig{file=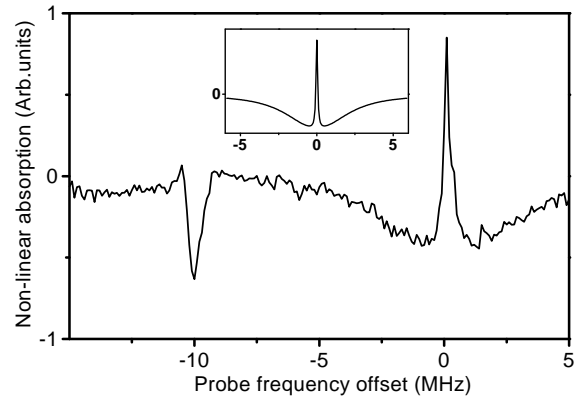,width=3.5in}}
\end{center}
\caption{Nonlinear absorption signal as a function of the pump to probe
frequency offset. Total trapping beam power $4\ mW$ (see text). Pump
frequency fixed at the center of the $5S_{1/2}(F=3)\rightarrow $ $%
5P_{3/2}(F^{\prime }=4)$ transition of $^{85}Rb$. Linear and orthogonal pump
and probe polarizations. Inset: calculated signal using the model in 
\protect\cite{LEZAMA3} for parameters $B=0$ and $\Omega _{1}=10^{-3}\Gamma $%
. }
\label{ancho}
\end{figure}

\begin{figure}[tbp]
\begin{center}
\mbox{\epsfig{file=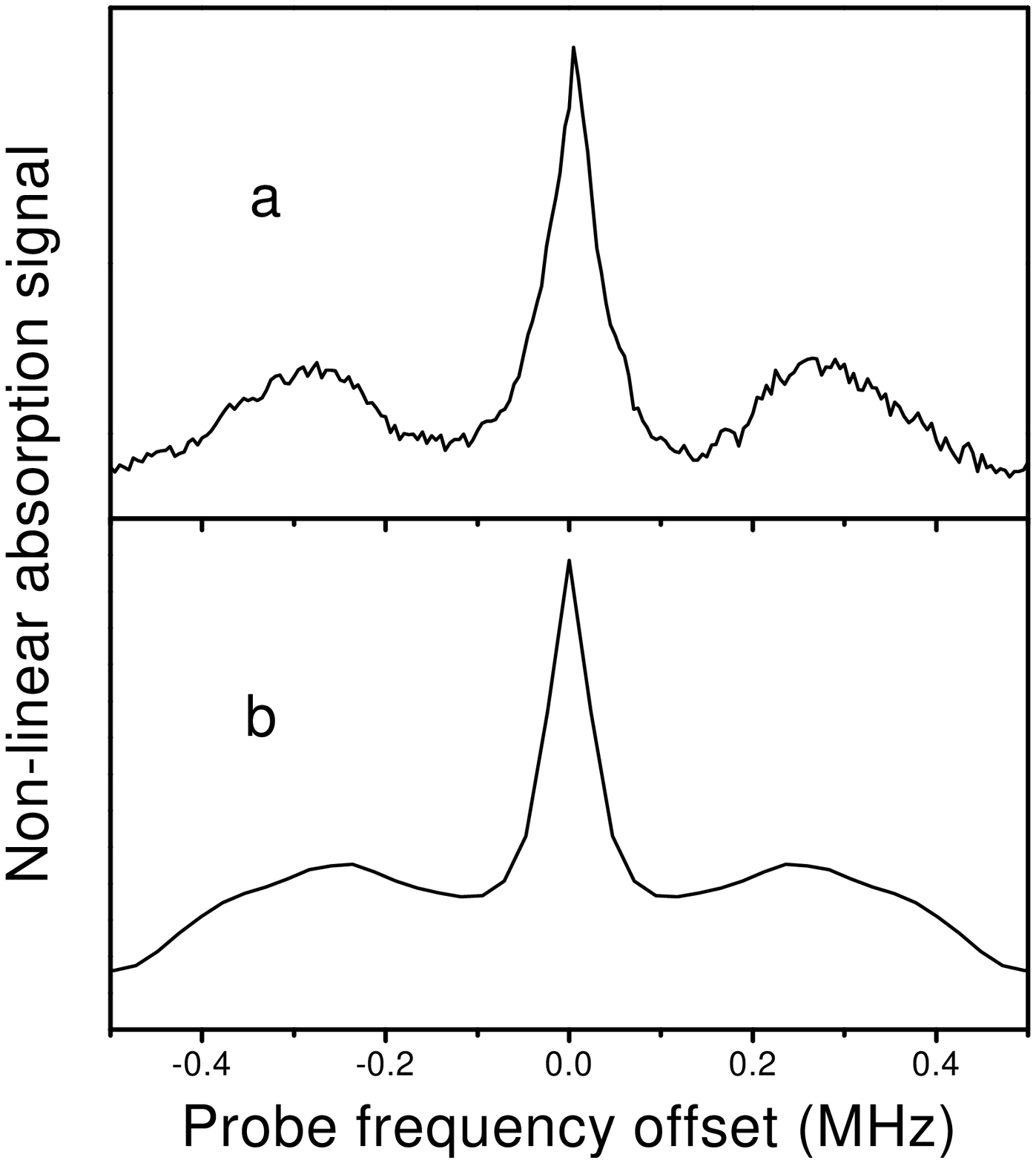,width=3.5in}}
\end{center}
\caption{a) Nonlinear absorption signal observed with total trapping beam
power $4\ mW$ (see text). The pump frequency is fixed at the center of the $%
5S_{1/2}(F=3)\rightarrow $ $5P_{3/2}(F^{\prime }=4)$ transition of $^{85}Rb$%
. b) Calculated nonlinear absorption spectra for a sample $1000$ atoms
randomly distributed in an oblate ellipsoidal volume centered in a
quadrupolar magnetic field with maximum value $B_{MAX}=0.8\ G$. The
calculation used the model in \protect\cite{LEZAMA3} with parameters: $%
\Omega _{1}=10^{-3}\Gamma ,$ $\protect\gamma =5\times 10^{-3}\Gamma $.}
\label{triplet2}
\end{figure}

An example of a low resolution nonlinear absorption spectrum as a function
of the frequency difference between the probe and the pump fields is
presented in Fig. \ref{ancho}. In this spectrum, the pump frequency is kept
fixed at the center of the $5S_{1/2}(F=3)\rightarrow $ $5P_{3/2}(F^{\prime
}=4)$ transition of $^{85}Rb$. The spectrum presents three distinct
features. Two of these are centered around $\delta =0$. There is a broad
resonance (dip) whose width is given by the excited state width $\Gamma
/2\pi =6MHz$. At the bottom of this resonance there is a much narrower one
with opposite sign. The broad resonance corresponds to an increase in the
transmission due to the saturation of the sample by the pump field while the
narrow feature represents an increase of the absorption. The latter
corresponds to EIA. The sub-natural width of this resonance is an indication
of its coherent nature. The inset in Fig. \ref{ancho} shows the calculated
signal around $\delta =0$ using the model described in \cite{LEZAMA2,LEZAMA3}%
. An interesting feature in the spectrum in Fig. \ref{ancho} is the large
dip appearing around $\delta \simeq -10\ MHz$. This resonance occurs when
the probe frequency equals the frequency of the trapping beams (red-detuned $%
10\ MHz$ with respect to the atomic transition). Since, as a consequence of
the technique used for the fields preparation, the probe, pump and trapping
beams are highly correlated, it is rather natural to observe a coherent
resonance involving the probe and the \ trapping fields. However, keeping in
mind the two-frequencies detection technique used, the pump field must also
participate in the generation of any detected signal. Consequently the peak
observed around $\delta \simeq -10\ MHz$ corresponds to a nonlinear process
involving at least one photon from each of the tree fields (pump, probe and
trapping). The study of this multiphoton effect is beyond the scope of this
paper. We focus our attention on the EIA structure observed around $\delta
=0 $. Fig. \ref{triplet2}(a) presents a detailed view of this structure
observed in a MOT with a total trapping beams power attenuated to
approximately $4m\ W/cm^{2}$. Under this condition the linear absorption by
the cold atoms was maximum ($40\%$) at the trapping transition. A triplet
structure is clearly visible which is reminiscent of the EIA spectrum
obtained in a homogeneous sample for perpendicular pump and probe
polarizations in the presence of a magnetic field orthogonal to both
polarizations \cite{LEZAMA3}. The observed value of the frequency separation
between the center and the two sidebands would correspond in that case to a
magnetic field of $B\approx 0.3\ G$. This figure is of the order of the
typical magnetic field magnitude within the cold atoms volume due to the
quadrupolar distribution. Since in our experiment the pump and probe beams
overlap a significant fraction ($20\%$) of the cold atomic sample, the EIA
spectrum is due to the average contribution from atoms in different magnetic
environments.

The hypothesis that the observed structure of the EIA spectrum is determined
by the magnetic field present at the MOT can be tested by adding a small
bias magnetic field at the sample. This was achieved by varying the current
in the Helmholtz coils surrounding the trap that are normally used to
compensate the earth field. The effect of the bias field is to change the
magnetic field distribution at the position explored by the pump and probe
waves. It also produces a translation of the cloud that results in a
variation in the total number of atoms interacting with the pump and probe
waves. Significant changes in the relative weight and positions of the
resonances occur for different bias magnetic fields. Since in these
measurements the pump, probe and trapping fields are kept unchanged, we
conclude that the observed spectral structure is mainly determined by the
magnetic field. When the pump-probe beam was carefully aligned through the
atomic cloud for maximum linear absorption, the observed structure of the
EIA spectrum was a triplet [Fig. \ref{triplet2}(a)]. This structure
represents the total response of the magnetically inhomogeneous sample. We
have compared this observation with the numerical calculation of the
response arising from $1000$ atoms in different positions in an ideal MOT.
We used a Monte Carlo procedure where the position of the atom is randomly
chosen (uniform distribution) within an oblate ellipsoidal volume,
representing the trapped cloud, centered at the zero of a quadrupolar
magnetic field. The vertical dimension (along the symmetry axis) of the
ellipsoid is shorter by a factor of two than the horizontal one. The
absolute value of the magnetic field at the points where the principal axis
intercept the ellipsoid surface is $B_{MAX}$. The pump field polarization is
vertical and the probe polarization horizontal. For each atomic position the
local magnetic field is evaluated \cite{SAVARD} and the corresponding
nonlinear response is calculated using the model described in \cite{LEZAMA3}%
. Then the total absorption is calculated as the sum of the individual
atomic contributions. The result of the simulation is shown in Fig. \ref
{triplet2}(b) where a realistic value of $B_{MAX}=0.8\ G$ has been chosen.
The total absorption spectrum is a triplet in agreement with our
observation. Unlike, the ideal ellipsoidal sample assumed for the
calculation, the actual cloud of cold atoms is not symmetric around the zero
of the magnetic field distribution. This is due to power imbalance and
wavefront irregularities in the counterpropagating trapping beams. This
produces a cold cloud displaced from the zero of the magnetic field with an
asymmetric shape and an irregular atomic density distribution. This explains
the fact that varying the position of the pump-probe beam across the atomic
sample we were able to observe significant variations of the nonlinear
absorption shape ranging from a single-peak spectrum, a doublet, a triplet
or even a five-peaks structure.

Considering that the EIA spectra were recorded with the trapping field on,
it is somehow surprising to observe spectral structures determined by the
Zeeman shift of ground-state sublevels by the quadrupolar magnetic field. In
the trapping region, the resonant interaction of the atoms with the trapping
field results in significant displacement of the atomic levels due to (AC
Stark) light shift. The magnitude of the light shift is different for
different ground-state sublevels and depends on the local trapping field
intensity and polarization experienced by the atom. The six trapping beams
are responsible for a rather complicate interference pattern resulting in
three dimensional modulation of the trapping field intensity and
polarization with a spatial period of half wavelength \cite{HOPKINS}. The
importance of the ground-state sublevels light shifts in a MOT is well
illustrated by the peculiar shape of the absorption spectra of the trapped
atoms in the MOT \cite{TABOSA} analyzed by Grison and coworkers \cite
{GRYNBERG}. The influence of the ground-state Zeeman sublevels light shifts
on the absorption spectra of strongly driven degenerate two-level systems
was studied in \cite{LIPSICH}. Since the light shifts are uncorrelated with
the Zeeman shift produced by the quadrupolar magnetic field, one should
expect that the former will result in an inhomogeneous dispersion of the
energies of the ground-state sublevels producing a smearing of the EIA
resonance frequencies.

\begin{figure}[tbp]
\begin{center}
\mbox{\epsfig{file=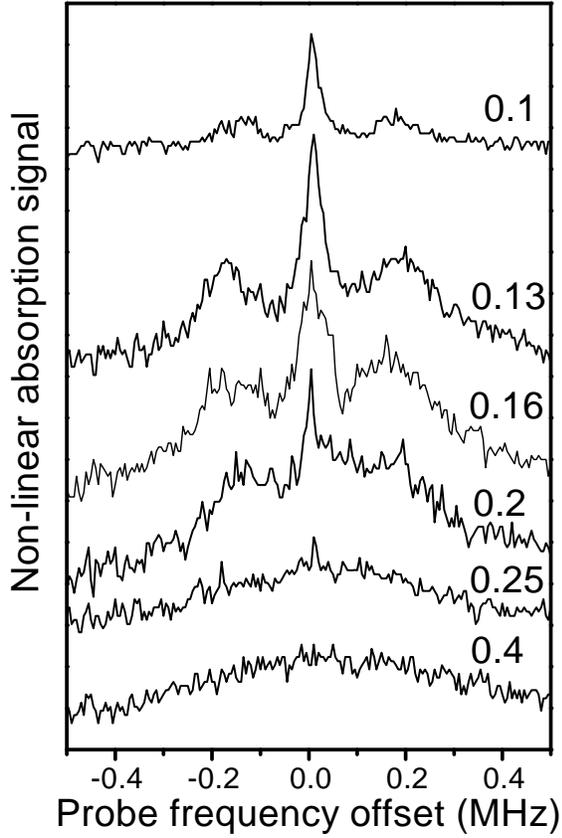,width=3.5in}}
\end{center}
\caption{Nonlinear absorption spectra for different intensities of the
trapping field. The relative trapping field power is indicated above each
trace. }
\label{trapintens}
\end{figure}

We have observed the variation of the EIA spectrum with the intensity of the
trapping field (Fig. \ref{trapintens}). For low trapping field intensities a
well resolved triplet was obtained. The width of the central peak of the
upper trace in Fig. \ref{trapintens} is $30\ kHz$. This is to our knowledge
the narrowest coherent resonance observed to date in an operating MOT \cite
{TABOSA,GRYNBERG,VELDT,HOPKINS2,DURRANT}. As the trapping field intensity is
increased, the EIA resonances broaden. For large enough trapping field
intensity, the triplet structure transforms into a unique broad peak. It is
worth noticing that a regime can be found where the trapped cloud is rather
dense (linear absorption around $40\%$) while at the same time the
magnetic-field-dependent structure of the EIA\ spectrum remains clearly
resolved. In this regime the modification of the ground-state sublevels by
the light shift produced by the trapping field is small compared to the
Zeeman shift.

\begin{figure}[tbp]
\begin{center}
\mbox{\epsfig{file=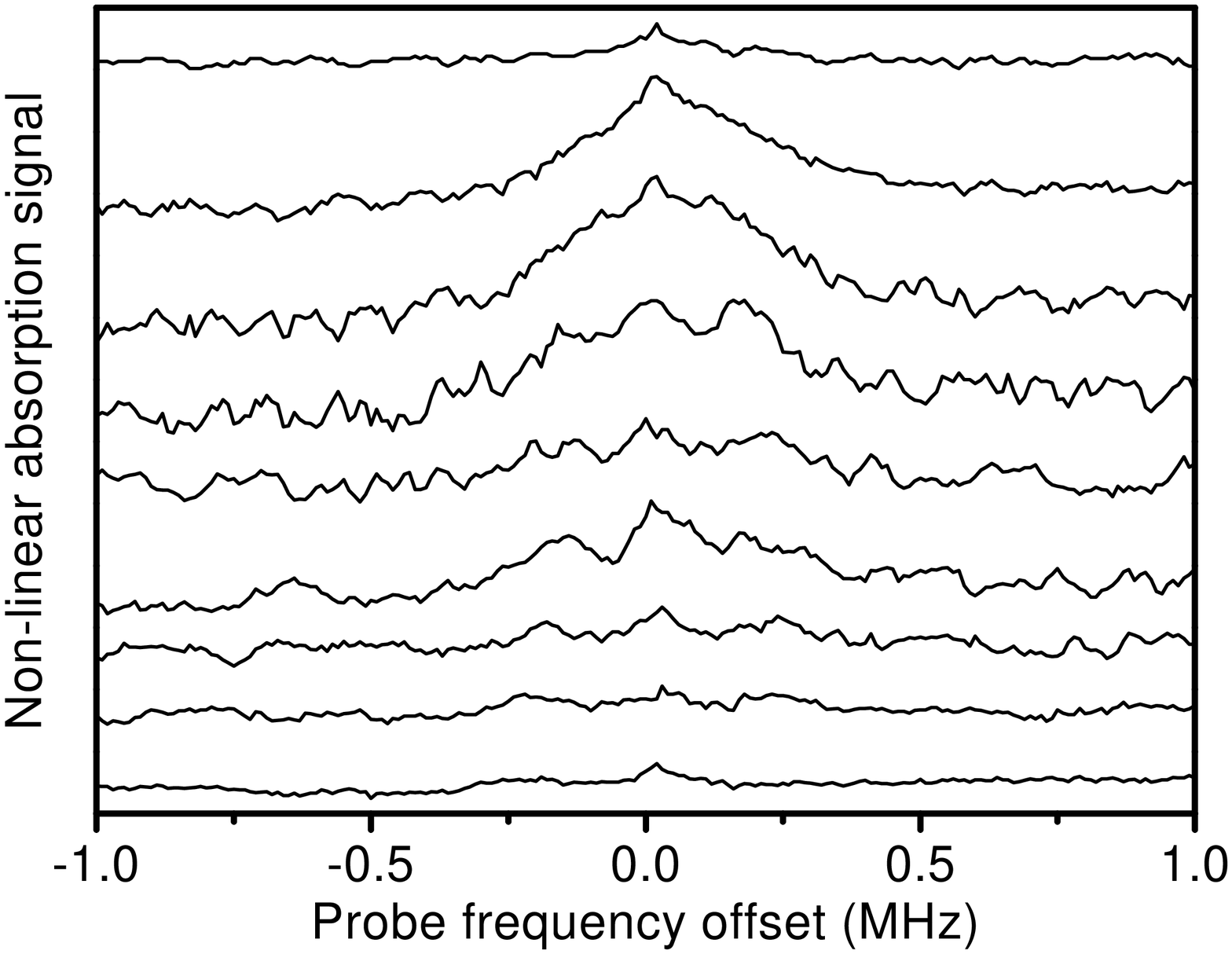,width=3.5in}}
\end{center}
\caption{Nonlinear absorption spectra obtained for different positions of
the pump-probe beam. Consecutive traces correspond to a $100\ \protect\mu m$
horizontal translation of the pump-probe beam.}
\label{tomografia}
\end{figure}

The fact that the EIA spectrum reflects the magnetic-field distribution
within the trapped atoms cloud may be used for mapping the magnetic-field
variations inside the cold atoms sample. This possibility is illustrated in
figure \ref{tomografia} showing the modification of the EIA spectrum as the
pump-probe beam is translated in the equatorial plane of the cold atom
cloud. Between two consecutive traces of Fig. \ref{tomografia} the
pump-probe beam was translated by $100\ \mu m$. From bottom to top, the
spectra in figure \ref{tomografia} evolve from a well resolved tripled into
a broad unresolved peak. The series in Fig. \ref{tomografia} may be
interpreted as a mapping of the magnetic field inside the cold atoms cloud
averaged over the volume explored by the pump-probe beam. In these
measurements the pump-probe beam diameter at the sample is $0.3\ mm$ a
figure which is only a factor of $\ 2.3$ smaller than the estimated cold
atoms cloud diameter. This limits in our case the spatial resolution of the
magnetic field mapping. A two orders of magnitude improvement in the
resolution is in principle possible with a tight focussed pump-probe beam.
The spectra presented in Fig. \ref{tomografia} clearly indicate the lack of
symmetry of our MOT. As a matter of fact, the zero of the magnetic field,
around which symmetric variations of the EIA spectrum should take place, is
located outside the region of large atomic density.

The mapping of the magnetic field using EIA spectroscopy, appear as a rather
simple and straightforward technique. It can be achieved in a cw regime
without the need for turning off the MOT trapping beams and with very
low-power pump and probe fields. As demonstrated here, these fields can be
easily obtained from the trapping laser. As in the example shown,
qualitative information about the atomic density distribution with respect
to the quadrupolar magnetic field is directly inferred from the spectra. The
fitting of the experimental EIA spectra to theoretical predictions (based on
consistent assumptions for the trapped cloud density distribution) can in
principle be used as a tomographic technique for the characterization of the
atoms in the MOT \cite{SAVARD}. This approach was not developed in the
present case in view of the irregular atomic density distribution observed
in our MOT.

\section{Conclusions.}

We have observed EIA resonances on a sample of cold atoms in a MOT in the
presence of the trapping light beams. An operating regime of the MOT (with
moderate trapping beams intensity) could be found where the atomic density
is quite large while at the same time the atomic ground state is weakly
perturbed by the trapping field. In this regime the energy shift of the
ground state sublevels is essentially governed by the Zeeman effect.
Coherent resonances as narrow as $30\ kHz$ were observed for the first time
in an operating MOT. EIA absorption spectroscopy, performed in a cw regime
with very low optical intensities, appears as a suitable tool for the
inspection of the Zeeman shift produced for the quadrupolar magnetic field
at different positions of the cold atom sample.

\section{Acknowledgments.}

The authors are thankful to A. Akulshin for his contribution to the early
stages of this study and to H. Failache for helpful discussions. This work
was supported by the Uruguayan agencies:\ CONICYT, CSIC and PEDECIBA and by
the ICTP/CLAF.


\end{document}